\documentclass[prb,twocolumn,superscriptaddress,longbibliography,aps]{revtex4-1}

\usepackage[T1]{fontenc}
\usepackage[utf8]{inputenc}
\usepackage{lmodern}
\usepackage[sumlimits]{amsmath}
\usepackage{amssymb}
\usepackage{bm}
\usepackage{yhmath}
\usepackage{graphicx}
\usepackage[no-test-for-array]{nicematrix}
\usepackage{epstopdf}
\usepackage{latexsym}
\usepackage{color}
\usepackage{nicefrac}
\usepackage{dsfont}
\usepackage{bbold}
\usepackage{wasysym} 
\usepackage{stmaryrd}
\usepackage{hyperref} 
\usepackage{xcolor}
\usepackage{multirow}
\usepackage{soul}
\usepackage{adjustbox}
\usepackage{cancel}

\usepackage{esint}
\usepackage{tikz}
\usepackage{mathtools}

\newcommand{\be}{\begin{equation}}
\newcommand{\ee}{\end{equation}}
\newcommand{\bea}{\begin{eqnarray}}
\newcommand{\eea}{\end{eqnarray}}

\usepackage{verbatim}
\usepackage{dsfont} 
\usepackage{caption}
\usepackage{url}

\usepackage{subcaption}

\usepackage{sansmath}
\DeclareFontEncoding{LGR}{}{}
\DeclareSymbolFont{sfgreek}{LGR}{cmss}{m}{n}
\SetSymbolFont{sfgreek}{bold}{LGR}{cmss}{bx}{n}
\DeclareMathSymbol{\sxi}{\mathord}{sfgreek}{`x}
\DeclareMathSymbol{\stheta}{\mathord}{sfgreek}{`j}
\DeclareMathSymbol{\sepsilon}{\mathord}{sfgreek}{`e}
\DeclareMathSymbol{\sOmega}{\mathalpha}{sfgreek}{`W}
\DeclareMathSymbol{\stau}{\mathalpha}{sfgreek}{`t}

\newcommand{\mb}{\mathbf}
\newcommand{\bs}{\boldsymbol}

\usepackage{scalerel}

\usepackage{accsupp}

\begin{document}

\title{Probing topology in thin films with quantum Sondheimer oscillations}

\author{L\'eo Mangeolle}
\affiliation{Technical University of Munich, TUM School of Natural Sciences, Physics Department, 85748 Garching, Germany}
\affiliation{Munich Center for Quantum Science and Technology (MCQST), Schellingstr. 4, 80799 M{\"u}nchen, Germany}
\author{Johannes Knolle}
\affiliation{Technical University of Munich, TUM School of Natural Sciences, Physics Department, 85748 Garching, Germany}
\affiliation{Munich Center for Quantum Science and Technology (MCQST), Schellingstr. 4, 80799 M{\"u}nchen, Germany}

\date{\today}
\begin{abstract}
Sondheimer oscillations (SO) are magnetoresistance oscillations occurring in thin films due to the commensurability between cyclotron motion and sample thickness, and are traditionally regarded as a purely semiclassical size effect. Here we develop a general quantum theory of SO for thin-film conductors in the quantum limit of a large magnetic field. We show that corrections arising from band topology modify the SO frequency, in contrast to Shubnikov–de Haas oscillations where topological information appears only in the phase. As a consequence, quantum SO provide a direct and robust probe of the full Landau level spectrum. Applying our framework to a minimal model with tunable Berry phase, we demonstrate how topology manifests itself in experimentally accessible magneto-oscillation spectra and  discuss  damping mechanisms including surface roughness. 
\end{abstract}

\maketitle

\section{Introduction}
\label{sec:introduction}

Magneto-oscillation phenomena describe the periodic variation of observables of conducting materials in applied magnetic fields $B$. The well-known Shubnikov de Haas (SdH) effect, succinctly described by Onsager \cite{onsager1952interpretation}, is a direct signature of the quantum mechanical quantization of energy levels in a magnetic field~\cite{landau1930diamagnetismus} leading to periodic modulation in $1/B$. A distinct phenomenon are Sondheimer oscillations (SO), which appear in thin films when the mean free path is at least of the order of the sample thickness. They were first predicted by Sondheimer for free electrons in a transverse field \cite{PhysRev.80.401,Sondheimer01011952}, then generalized within a semiclassical Boltzmann treatment to other dispersions and field orientations \cite{gurevich1959oscillations}. The basic semiclassical picture is that the spiral trajectories of charge carriers in a transverse $B$ field consist of an out-of-plane bouncing motion between the film boundaries, and an in-plane cyclotron motion (drifting in the presence of an electric field). Then SO arise, purely classically, from a simple commensurability condition between these two periodic components of the motion. \footnote{Helpful sketches of this mechanism, which is not our focus here, can be found in the literature.}

The theoretical description of SO and their early measurements in elemental metals~\cite{grenier1966magnetic,PhysRevB.8.5567,PhysRev.172.718,PhysRevLett.23.1287,PhysRevB.1.2385,trodahl1971sondheimer,sakamoto1976galvanomagnetic,alstadheim1968sondheimer,sato1979sondheimer} date back multiple decades, but they have remained little studied. They have been thought to be less useful for determining the electronic structure of metals compared to their widely used brethren SdH oscillations. Some theoretical puzzles they posed were only addressed recently \cite{nikolaenko2025theorysondheimermagnetooscillationssemiclassical}, and their experimental usefulness has long remained somewhat basic, in determining $k_z$-dependent features of the electronic bandstructure -- still with interesting implications e.g.\ investigating surface modes \cite{PhysRevB.84.125144} or very recently the Yamaji effect \cite{nikolaenko2026sondheimer}. SO have now received renewed interest thanks to the development of increasingly clean 2D nanostructures \cite{PhysRevB.108.235411,van2021sondheimer,PhysRevMaterials.9.014005}, where they were proposed in particular as indicator of non-Ohmic electron flow, and for the possibility to reach the quantum (large $B$) limit \cite{PhysRevB.108.235411}. In this work, we develop a theory of fully quantum SO (due to wavevector quantization in the transverse direction) in the quantum limit, and discover that they can also be used as a direct probe of \emph{band topology}.

The effects of band quantum geometry (including topology) in conventional magneto-oscillations à la Onsager are by now well established~\cite{alexandradinata2023fermiology}. It is possible to include the geometrical Berry phase \cite{roth1966semiclassical,wilkinson1984example} in the semiclassical quantization of electron orbits \cite{onsager1952interpretation}, which determines the semiclassical quantization of Landau levels \cite{fuchs2010topological}, and in turn SdH measurements. There, band-topological effects appear in the \emph{phase} of oscillations: for instance, the nontrivial winding number in graphene induces a dephasing of $\pi$. Such dephasing can be accessed experimentally via Landau level index plots (for a review see Ref.\cite{Zhao31122022}), but their precise determination is challenging, as it usually requires an extrapolation to $1/B \rightarrow 0$. Moreover, other dephasing effects can contribute, induced e.g. by interactions \cite{wasserman1996influence} or by dimensionality -- notably, even within standard Lifshitz-Kosevich theory \cite{lifshitz1956theory} a dephasing of $\pm \pi/4$ exists between 2D and 3D dispersions. This and other bandstructure effects can obscure signatures of topology from SdH experiments~\cite{fuchs2010topological,wright2013quantum,fuchs2018landau}.

Here, we propose to use {\it quantum SO}, found in the quantum regime of large magnetic fields and mean free paths exceeding the film thickness, to directly access information about band topology, avoiding the aforementioned issues. Indeed, we show that the energy level shifts induced by band topology do not appear in the phase of oscillations, but directly in their \emph{frequency}. Its determination is thus immune to spurious dephasing effects and does not require any interpolation. We show this by developing a theory of quantum SO that holds generally for any quasi-2D electronic Hamiltonian in a magnetic field. However, to motivate and illustrate our theory we apply it to a concrete tuneable model Hamiltonian $H_\lambda, \,\lambda \in [0,1]$, such that $H_0$ and $H_1$ have the exact same zero-field dispersion but different LL spectra because of their distinct topology. We note that a similar family of Hamiltonians was investigated in the context of the $\alpha-T_3$ model \cite{PhysRevLett.112.026402, PhysRevB.92.245410}. Moreover, all our $H_\lambda$ display the exact same SdH dephasing, but we show their different topological properties can be distinguished from the quantum SO spectra.

\begin{figure*}[!t]
\centering
\includegraphics[width=\textwidth]{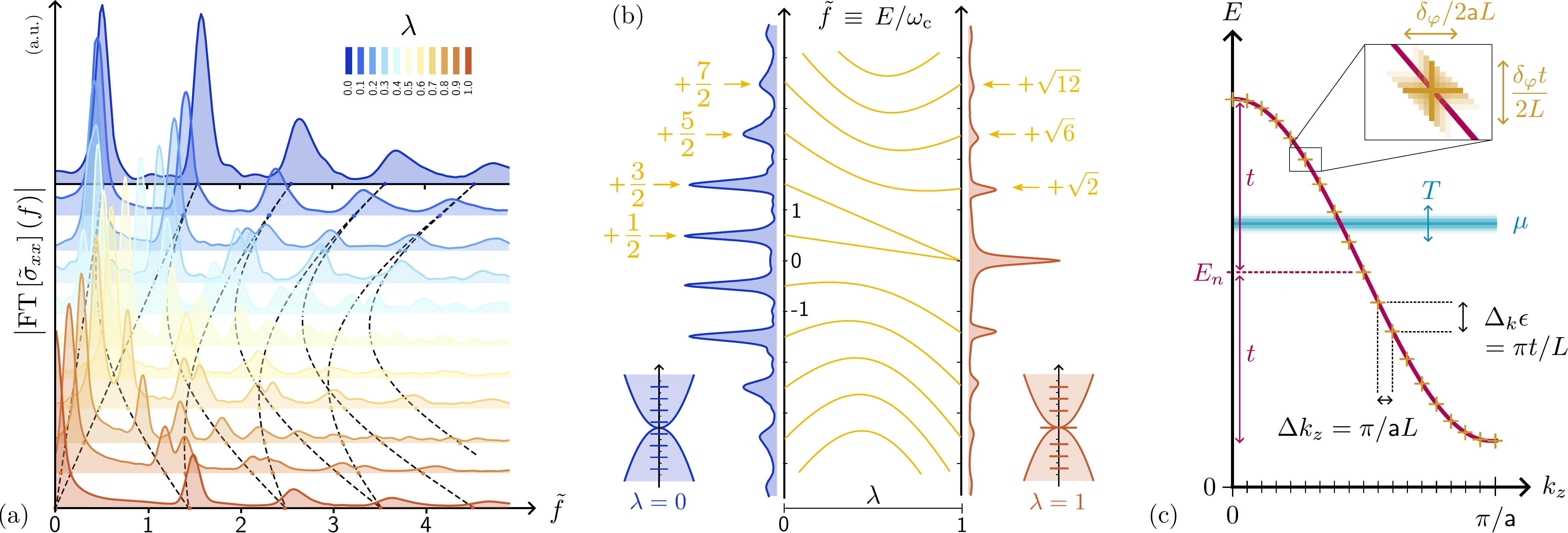}
\caption{\textbf{(a)} Fourier spectrum of quantum SO given by Eq.\eqref{eq:54} at $T=0$ for different values of $\lambda$, as a function of $\tilde f = (tm/2{\sf e}L)\, f$, defined from $\sigma_{xx}(B) = \int \text d f \,e^{-i f B} \,{\rm FT}[\sigma_{xx}](f)$. 
  Dashed black lines: LL spectrum of $H_\lambda$, folded to positive energies, with the same axes, identifying $\tilde f \equiv E/\omega_{\rm c}$.
  \textbf{(b)} Yellow: LL spectrum of $H_\lambda$ as a function of $\lambda \in [0,1]$, and special values at $\lambda=0,1$.
  Blue, red (for $\lambda=0,1$): reproduction of ${\rm FT}[\tilde \sigma_{xx}](f)$, symmetrized to negative energies;
  insets: sketch of the dispersion and LL spectrum in both limits. \textbf{(c)} Summary of the mechanism for quantum SO.
  For a given LL $n$, when $B$ (thus $E_n$) is varied, the whole dispersion shifts vertically, so discretized energy levels (brown reticles) periodically cross the Fermi level (where $v_z=t/\sf a$).
Both thermal broadening (teal) and surface roughness (brown) generate damping of SO.}
\label{fig:bigfigure}
\end{figure*}

\section{Setup}
\label{sec:setup}

We start from two different Hamiltonians for a quadratic band touching of 2D electrons:
\begin{subequations}
  \begin{align}
  \label{eq:1a}
    H_0 &= (p_x^2+p_y^2) \hat \sigma^z /2m ,\\
    \label{eq:1b}
    H_1& =\left [ (p_x^2-p_y^2)\hat \sigma^x + 2 p_xp_y \hat \sigma^y \right ] /2m ,
         \end{align}
     \end{subequations}
     where $\hat \sigma^x, \hat \sigma^y, \hat \sigma^z$ are the Pauli matrices.
Both Hamiltonians have exactly the same dispersion $\epsilon^\pm_{\bs p} = \pm p^2/2m$, however they are topologically distinct.
Indeed $H_0$ describes a topologically trivial, i.e. fine-tuned, quadratic band touching \footnote{This is realized near any trivial gap in semiconductors, since the bands' actual touching is irrelevant in the present discussion.},
and its spectrum in a magnetic field $B$ is simply $E_{0,n}^\pm = \pm \omega_{\rm c}(n+1/2), n\in \mathbb N$.
Meanwhile $H_1$ describes a topologically nontrivial quadratic band touching as in AB-stacked bilayer graphene \cite{mccann2006landau,novoselov2006unconventional},
and its spectrum in a magnetic field has topologically protected zero-modes, $E_{1,n}^\pm = \pm \omega_{\rm c} \sqrt{n(n+1)}, n\in \mathbb N$.
In both cases $ \omega_{\rm c}={\sf e}B/m$. In the semiclassical limit of $n\rightarrow \infty$ (equivalently, in zero field) these spectra are identical.

Let us define $  H_\lambda = \lambda   H_{1} + (1-\lambda)  H_0 $, for $\lambda \in [0,1]$, that interpolates between these two Hamiltonians.
It is soluble in a field for all $\lambda$, but first it is instructive to look at its semiclassical ($n\rightarrow \infty$) limit.
The semiclassical cyclotron orbits, enclosing an area $S_{\rm c}$ in momentum space,
are constrained by Roth's quantization formula, $l_B^2 S_{\rm c} = 2\pi (n+\tfrac 1 2 + \gamma_{\rm R})$,
where $n \in \mathbb Z$ and $\gamma_{\rm R}$ is minus the Berry phase accumulated around the orbit \cite{roth1966semiclassical,wilkinson1984example}.
Here, say for the upper band, one finds (see appendix \ref{sec:few-notes-about}) $1-\gamma_{\rm R}(\lambda) =  (1-\lambda)/\sqrt{\lambda^2 + (1-\lambda)^2}$,
which does not distinguish between the two Hamiltonians, $\gamma_{\rm R}(0) = \gamma_{\rm R}(1) \;\rm mod\,1$.
In fact, it is known \cite{PhysRevB.77.245413,fuchs2010topological} that for such two-band models, the semiclassical Landau level (LL) energies do not depend directly on $\gamma_{\rm R}$ but only on its topological part,
$\gamma_{\rm L}=- w_{\rm c}/2$ where $ w_{\rm c}$ is the winding number of the cyclotron orbit.
Here, $-w_{\rm c}(\lambda)=2(1-\delta_{\lambda,0})$ which does not allow to distinguish between \emph{any} values of $\lambda \in [0,1]$. \cite{PhysRevLett.112.026402, PhysRevB.92.245410}

Next, we obtain the exact LL spectrum of $H_\lambda$ in a field. It consists of a symmetrical set of ``regular'' levels (indexed as $|\tilde n,s\rangle,\tilde n\geq 2,s=\pm$) and two zero modes ($|0,\uparrow\rangle$ and $|1,\uparrow\rangle$).
Their energies are plotted as a function of $\lambda$ in the central part of Fig.\ref{fig:bigfigure}(b) -- the zero modes correspond to the two straight lines,
and pairs of modes $|n,\pm\rangle$ have the same energy up to a sign at $\lambda=1$.
Their expressions are  $E_0^\uparrow = +\tfrac 1 2 \omega_{\rm c}(1-\lambda) , E_1^\uparrow = +\tfrac 3 2 \omega_{\rm c}(1-\lambda)$ and
$ E_{\tilde n}^\pm =  \omega_{\rm c} \left [ (1-\lambda) \;\pm\; \sqrt{(1-\lambda)^2 (\tilde n-\tfrac 1 2)^2+\lambda^2 \tilde n(\tilde n-1)} \right ]$ -- for details, see appendix \ref{sec:land-levels-energ}.
For later convenience let us define a new index $n \in \mathbb Z$ to label LLs, so that $\sum_n {\rm f}_n := \sum_{\tilde n\geq 2,s=\pm}  {\rm f}_{\tilde n,s} + \sum_{\eta=0,1} {\rm f}_{\eta,\uparrow}$ for any $\rm f$,
and the corresponding energies $E_n$ \footnote{For instance one can choose $|\tilde n,\pm\rangle \mapsto n=\pm \tilde n$ and $|\eta,\uparrow \rangle \mapsto n=2\eta-1$, so that $n$ spans $\mathbb Z - \{0\}$}. 
It is clear from Fig.\ref{fig:bigfigure}(b) that the low-energy part of the spectrum (i.e. $|E_n| \lesssim \omega_{\rm c}$, the region poorly captured by semiclassics)
allows one to distinguish between $H_0$ and $H_1$, and more generally all $H_\lambda$.
It is the purpose of our work to show how these can be efficiently extracted from the frequency of SO in this quantum regime.

\section{Quantum Sondheimer oscillations}
\label{sec:quant-sondh-oscill}

\subsection{General theory}
\label{sec:general-th}

To do this, we now consider a slab geometry, with $L$ two-dimensional layers indexed by $l\in \llbracket 1,L\rrbracket$
where the Hamiltonian is $H_{\rm layer}$ (in our case, we will later take $H_{\lambda}$). The total three-dimensional Hamiltonian is
\begin{align}
  \label{eq:24}
  H_{\rm tot}= H_{\rm layer}  - t \sum_{l=1}^{L-1}\int_{x,y}\left ( \psi_{l}^\dagger \psi_{l+1}+ \psi_{l+1}^\dagger \psi_{l} \right ) (x,y),
\end{align}
where $ \psi_{l}^\dagger (x,y)$ is the fermion creation operator in layer $l$ at position $(x,y)$ and $t$ is the unit of kinetic energy.
This finite geometry quantizes momentum along the $z$ axis, with electron wavefunctions proportional to $\sin(\pi k/L)$ associated with a kinetic energy $-t\cos(\pi k/L)$
\footnote{Strictly speaking the argument should be $\pi k/[L+1]$ for both, but to avoid unnecessary cluttering we write $\pi k/L$ instead throughout the paper.},
indexed by $k \in \llbracket 1,L\rrbracket$. This discretization, seen in experiments \cite{yin2019dimensional}, will be responsible for SO.
The energy for each pair of indices $(n,k)$ is
\begin{align}
  \label{eq:43}
  E_{n,k}&= E_n -t\cos(\pi k/L) ,
\end{align}
and each such level has a degeneracy indexed by momentum ${\rm p}_x \in \llbracket 0 ; {2\pi}/L_x ; \cdots ; L_y l_B^{-2} \rrbracket $,
yielding the usual LL degeneracy $N_{\sf \Phi} = L_x L_y/2\pi l_B^2$ ~\cite{landau1930diamagnetismus}.
We stress that the whole discussion holds regardless of the structure of $E_n$ and holds for any two-dimensional Hamiltonian $H_{\rm layer}$:
all the relevant features, for a given level $n$, are summarized in Fig.\ref{fig:bigfigure}(c).

We now consider the conductivity kernel \cite{bastin1971quantum}
\begin{align}
  \label{eq:27}
   \sigma_{xx}(\mu) &= - \frac{\sf e^2}{2\pi^2} \frac{l_B^{-2}}{{\sf a}L}\, {\rm Tr}_{k,n}\left [ \hat v_x \, {\rm Im}G(\mu) \,\hat v_x \, {\rm Im}G(\mu) \right ] ,
\end{align}
where $\sf a$ is the interlayer spacing and the trivial sum over ${\rm p}_x$ is already taken. 
The electron propagator is $G_{n,k}(\mu) = \left [ \mu - E_{n,k} + i\Gamma_n \right ]^{-1}$, diagonal in the $(n,k)$ basis. Note, we assume that the effect of disorder can be captured by a simple damping rate $\Gamma_n$ (that we assume to be independent of $k$). We neglect any oscillatory dependence of $\Gamma_n$ on the magnetic field~\footnote{Such oscillations can yield rich physics\cite{mangeolle2025anomalous}, but make the whole derivation more cumbersome and are not expected to change the lowest harmonics of oscillations. Neglecting them is justified in many limits \cite{wasserman1996influence}.}.
We also neglect vertex corrections from disorder, so the velocity operator $\hat v_x=\partial H_\lambda / \partial p_x$ is the bare one.
Its matrix elements connect different values of $n$ but not of $k$, and are $k$-independent: for details about $ \langle n| \hat v_x | n' \rangle$, see appendix \ref{sec:velocity-operator}.

To derive magneto-oscillations from a formula such as Eq.\eqref{eq:27}, a standard way is to use Poisson resummation.
While Onsager (including SdH) oscillations require resumming over $n$, quantum Sondheimer oscillations arise from resumming over $k$, using
$ \sum_k {\rm g}(k) = \sum_{r \in \mathbb Z} \int\text dk \, e^{i2\pi r k} \,{\rm g}(k) $ for any $\rm g$.
We focus on the $B$-periodic oscillatory part, so we discard the $r=0$ term ; the latter still contains the $1/B$-periodic SdH oscillations (see appendix \ref{sec:sdh}), and is thus crucial for understanding the crossover between the SdH and quantum Sondheimer regimes, which goes beyond the scope of this study. In addition, we keep only the leading harmonics, namely $r=\pm 1$ (justified later).
In the following this is denoted $\tilde  \sigma_{xx}(\mu) $.
Then, the $\int\text dk $ can be computed exactly using contour integration (cf appendix \ref{sec:techn-deta-summ}).
The oscillation frequencies are determined by the pole $k^\star_{n} =  L \arccos[ (E_n - \mu - i \Gamma_n)/t] /\pi$.
One eventually obtains
\begin{align}
  \label{eq:54}
   \tilde \sigma_{xx}(\mu) &= - \frac{\sf e^2}{\pi l_B^2} \left ( \frac{L}{\pi t} \right )^2  \frac 1 {{\sf a}L} \sum_n 
    2 \, {\rm Im} \left ( e^{ i2\pi k^\star_{n}} K_n \right ),\\
 K_{n} = \sum_{n'} & \frac{\big | \langle n| \hat v_x | n' \rangle \big |^2}{ \sin(\pi k^\star_{n}/L)}\,
                     \Big ( \frac { \left [ k^\star_{n}- k^\star_{n'} \right ]^{-1} } { \sin(\pi k^\star_{n'}/L) }
                     - \frac { \left [  k^\star_{n}-\bar k^\star_{n'} \right ]^{-1} }{ \sin(\pi \bar k^\star_{n'}/L) } \Big ) , \nonumber
\end{align}
where $\bar k^\star_{n} = (k^\star_{n})^*$.
One can show (see appendix \ref{sec:techn-deta-summ}) that the double sum $\sum_{n,n'}$ indeed converges,
as it selects contributions from levels where $ E_{n}$ is reasonably close to the Fermi level.

At this stage, Eq.\eqref{eq:54} is our general result for quantum SO derived from a purely quantum-mechanical calculation starting from Eq.\eqref{eq:24},
that does not rely on a specific in-layer Hamiltonian $H_{\rm layer}$. Now we come back to our original motivation that is $H_{\rm layer} \rightarrow H_{\lambda}$.

\subsection{Spectral flow from direct FT}
\label{sec:spectral-flow-from}

Before pursuing further analytical understanding, we simply evaluate $\tilde \sigma_{xx}(\mu)$ numerically, for the case of $\mu=0$ and of a single relaxation rate $\Gamma_n \rightarrow \Gamma$. 
As one would do experimentally, we then plot $\tilde \sigma_{xx}$ as a function of the magnetic field $B$ (not $1/B$ as one would do for Onsager oscillations) in a given range of $B$
-- the parameters we used are $m=1, L=200, t=4.0, \Gamma = 0.2, B \in [0.1, 2.1]$.
We then compute the Fourier transform, defined by $\tilde \sigma_{xx}(B) = \int \text d f \,e^{-i f B} \,{\rm FT}[\tilde \sigma_{xx}](f)$.
The result of this computation is displayed in Fig.\ref{fig:bigfigure}(a), for different values of $\lambda \in [0,1]$.
The particular cases $\lambda=0,1$ are also reproduced along the vertical axes of Fig.\ref{fig:bigfigure}(b).

Remarkably, the peaks of ${\rm FT}[\tilde \sigma_{xx}]$ are in one-to-one correspondence with the spectrum of LLs of $H_{\lambda}$.
To highlight this, in Fig. \ref{fig:bigfigure}(a) we plot as black dashed lines the lowest few energies $E_n(\lambda)$, folded to positive frequencies:
they track the positions and degeneracies of the dominant peaks.
In other words, this shows that a simple Fourier analysis of SO allows to track the spectral flow of $H_\lambda$,
by extracting the LL energies $E_n = \omega_{\rm c}\tilde f_n$ for each peak labeled by $n$ and with the rescaling $\tilde f = (tm/2{\sf e}L)\, f$.
In particular, it allows to distinguish unambiguously $H_1$ from $H_0$, as is evidenced in Fig.\ref{fig:bigfigure}(b):
the dominant peak is at $\tilde f = 1/2$ for $H_0$ and at $\tilde f=\sqrt 2$ for $H_1$. This constitutes the main finding of our work: for SO the topology of the band structure, e.g. the presence of zero-modes in the resulting Landau levels, is encoded directly in the \emph{frequency} of magneto-oscillations.

\section{Amplitude of the oscillations}
\label{sec:amplitude-osc}

\subsection{Damping factors}
\label{sec:thermal-damping}

Next, we explain analytically the numerical findings.
For energy levels located around $\mu \approx 0$, one can Taylor-expand the pole $k^\star_n$ so the oscillatory factor in Eq.\eqref{eq:54} is
\begin{align}
  \label{eq:1}
 e^{i2\pi k_n^\star} &\simeq  (-1)^L\,e^{-2\Gamma L/t}\,e^{-i2(E_n-\mu)L/t } .
\end{align}
Because the LL energies are proportional to the magnetic field, $E_n=\omega_{\rm c} \tilde f_n$ where $\omega_{\rm c}={\sf e}B/m$,
the last factor in Eq.\eqref{eq:1} is $e^{-i f_n B}$ where $f_n = (2{\sf e}L/tm) \tilde f_n$.
This explains the correspondence between the Landau spectrum $E_n(\lambda)$ and peaks in the Fourier transform of $\tilde \sigma_{xx}$ with respect to $B$. 
Note that $K_n$ in Eq.\eqref{eq:54} also depends on $B$, but on much slower scales (notice the $1/L$ in the oscillating factors), so it is not necessary to understand the oscillation spectrum. 
Besides, from Eq.\eqref{eq:1} one can notice that the phase of quantum SO depends on the precise value of the chemical potential $\mu$. 

\begin{figure}[!t]
\centering
\includegraphics[width=\columnwidth]{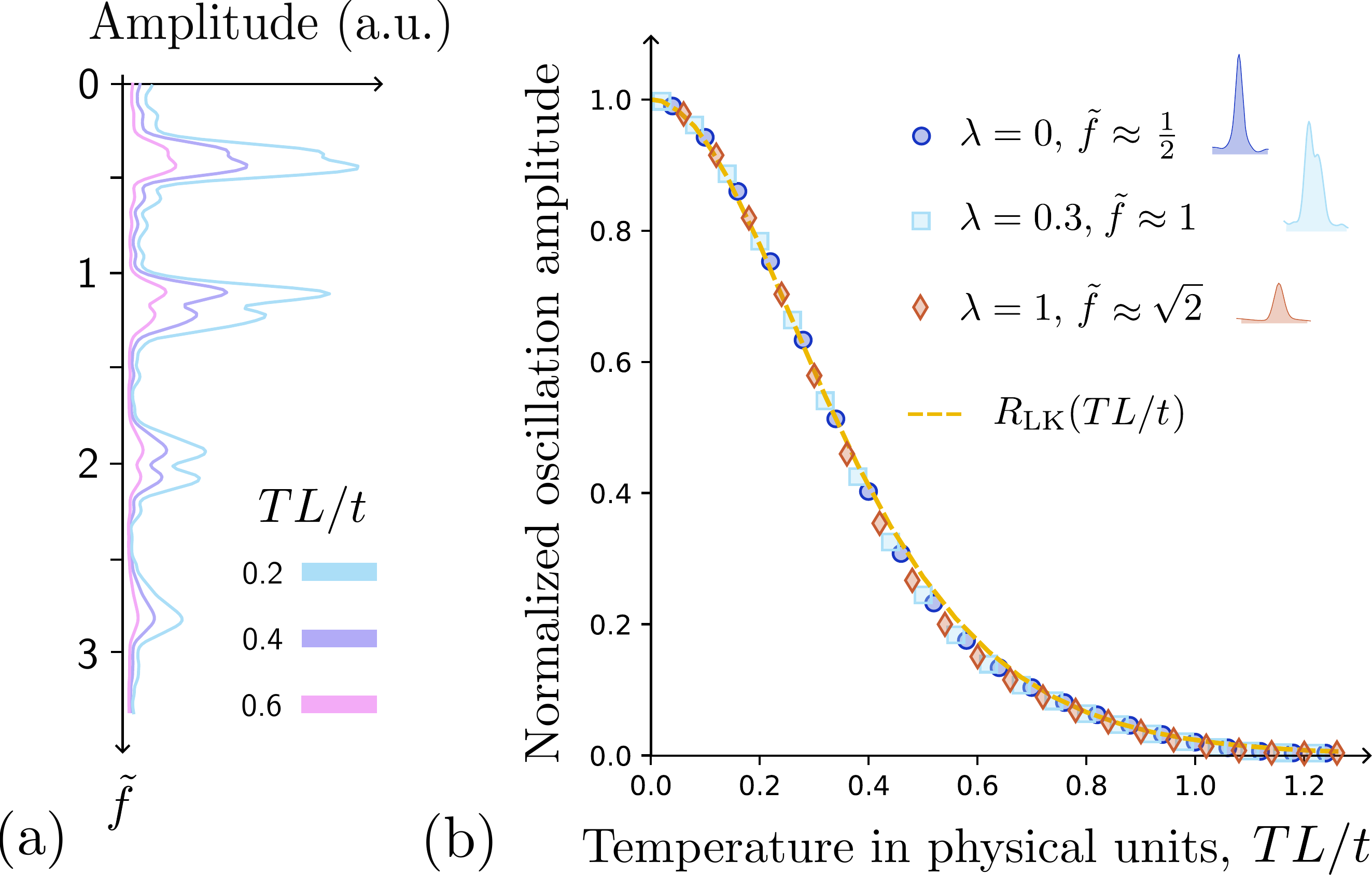}
\caption{\textbf{(a)} Fourier spectrum of quantum SO at $\lambda=0.3$ for different temperatures $T$,
  obtained directly from Eq.\eqref{eq:56} (``=''), as a function of $\tilde f = (tm/2{\sf e}L)\, f$.
  \textbf{(b)} Temperature dependence of the amplitude of three different peaks (at different $\lambda$),
  showing universal behavior matching the analytical result Eq.\eqref{eq:56} (``$\simeq$'').}
\label{fig:damping}
\end{figure}

Thus far, we have computed the zero-temperature oscillatory conductivity $\tilde \sigma_{xx}(\mu) =: \tilde \sigma^{(0)}_{xx}(\mu)$,
now we compute its finite-temperature counterpart by the standard convolution with the thermal broadening factor for fermions \cite{shoenberg1984magnetic} 
\begin{align}
  \label{eq:56}
  \tilde \sigma_{xx}^{(T)}(\mu) & =  \int \text d \epsilon \; \frac{ \tilde \sigma^{(0)}_{xx}(\mu+\epsilon) }{ 4T \cosh^2 \left ( \frac{\epsilon}{2T} \right ) }
                                  \simeq \tilde \sigma_{xx}(\mu)\,  R_{\rm LK}(LT/t),
\end{align}
where ``$\simeq$'' above follows from Eq.\eqref{eq:1} and a straightforward integration, and we use the shorthand $R_{\rm LK}(\chi) = 2\pi\chi / \sinh(2\pi \chi)$.
The validity of this approximation can be checked directly against the exact numerical evaluation.
To do this, we evaluate $\tilde \sigma_{xx}^{(T)}(\mu)$ using the convolution formula Eq.\eqref{eq:56}(left) together with the exact input Eq.\eqref{eq:54}.
Then we compute ${\rm FT}[\tilde \sigma^{(T)}_{xx}]$, the temperature dependence of which is illustrated in Fig.\ref{fig:damping}(a).
We then integrate over some given frequency intervals to obtain the total spectral weight of the dominant peak(s) for several values of $\lambda$,
which we plot as a function of $T$. The result is displayed in Fig.\ref{fig:damping}(b), which shows that the analytical result Eq.\eqref{eq:56}(right) is indeed very accurate.

We note that this thermal damping behavior as $ R_{\rm LK}(LT/t)$ is different from the usual Lifshitz-Kosevich dependence, $R_{\rm LK}(\pi T/\omega_{\rm c})$. Similarly, back to Eq.\eqref{eq:1}, the quasiparticle-lifetime broadening factor $e^{-2\Gamma L/t}$ is different from the usual Dingle factor $R_{\rm D} = e^{-2\pi \Gamma/\omega_{\rm c}}$. 
For both, the difference amounts to replacing $2\pi/\omega_{\rm c} \leftrightarrow 2L/t$. Identifying the velocity at the Fermi level $v_z={\sf a}t$,
this means the cyclotron period is replaced with $2 {\sf a}L/v_z$, the time it takes an electron to travel across the slab back and forth.
It is interesting to note that although the semiclassical interpretation is straightforward, this arises from a purely quantum-mechanical calculation,
and the quantization of kinetic energy along $\hat z$ with steps $\pi t/L$ instead of $\omega_{\rm c}$: see the sketch in Fig.\ref{fig:bigfigure}(c).

\subsection{Surface roughness}
\label{sec:surface-roughness}

Beside these damping effects due to thermal fluctuations and loss of quasiparticle coherence, that are present in any setting,
there are other damping effects that are specific for thin films, caused in particular by surface roughess and local strains.
In particular, we note that surface roughness affects the quantization of momentum $k_z$.
Indeed, the latter can be derived from a Bohr-Sommerfeld quantization condition,
\begin{align}
  \label{eq:60}
  2{\sf a}L k_z + \varphi \in 2\pi \mathbb Z ,
\end{align}
where $2{\sf a}L$ is the distance traveled by an electron crossing the slab back and forth, and $\varphi$ is the phase acquired at the boundaries.
Now crucially, $\varphi$ depends on the precise physical modelling of electron reflection at the surfaces of the film, and is affected by surface roughness.
One can model it \footnote{More detailed treatments of surface disorder and boundary scattering, in particular close to the semiclassical limit, can be found in Refs. \cite{PhysRevB.51.7325,nikolaenko2025theorysondheimermagnetooscillationssemiclassical}.} as a random variable with a probability distribution, which for simplicity we will take to be gaussian,
${\sf p}(\varphi)=(2\pi \delta_{\varphi})^{-1/2}\exp[-\varphi^2/2\delta_\varphi^2]$ with standard deviation $\delta_\varphi$.
This randomly shifts the discrete values of momentum $k_z$ allowed, and thus the energies as well, as depicted in the inset of Fig. \ref{fig:bigfigure}(c).
Averaging $e^{i2\pi k_n^\star}$ over $\varphi$ fluctuations in a similar fashion as Eq.\eqref{eq:56} yields an extra damping factor,
$R_{\Sigma} = {\rm FT}[{\sf p}](1)$, which in particular for the gaussian distribution we chose is $R_{\Sigma} = \exp \left (-\delta_{\varphi}^2/2\right ) $.
We note that Eq.\eqref{eq:60} fixes the momentum discretization step $\Delta k_z=\pi/{\sf a}L$ regardless of the precise open boundary conditions,
and holds provided that the mean free path is much longer than the film thickness, $\ell \gg {\sf a}L$ (equivalently $\Gamma L/t \ll 1$), which is the relevant limit for SO.

\subsection{Amplitude of higher harmonics}
\label{sec:temp-dep-harm}

Higher harmonics, $r E_n$ for a given LL energy $E_n$ and $r \in \mathbb Z, |r|\geq 2$, are suppressed by a Dingle-like factor $e^{-2|r|\Gamma/t}$, a thermal factor $ R_{\rm LK}(rLT/t)$, and especially by a surface damping factor $\exp\left (-r^2 \delta_{\varphi}^2/2\right )$. In the thin limit $\Gamma L/t \ll 1$, the latter are the main cause of harmonics suppression at low $T$. 

Their distinct temperature behaviors make it in principle easy to distinguish higher harmonics from the fundamental peaks of higher LL energies $E_{n'}, n' > n$. To illustrate this, we consider the case $\lambda = 0$, where we have restored all harmonics $r \neq 0$ in Eq.\eqref{eq:54} (see derivation in Appendix \ref{sec:techn-deta-summ}). We fixed the Dingle-like factor $R_{\rm D}\rightarrow 1$ and did not introduce any surface roughness, $\delta_\varphi =0$. Then, peaks are present at all $\tilde f_n = n+\tfrac 1 2, n\in \mathbb N$, and at all their harmonics, $r \tilde f_n, \,r=2,3,\cdots$. 

\begin{figure}[!t]
\centering
\includegraphics[width=\columnwidth]{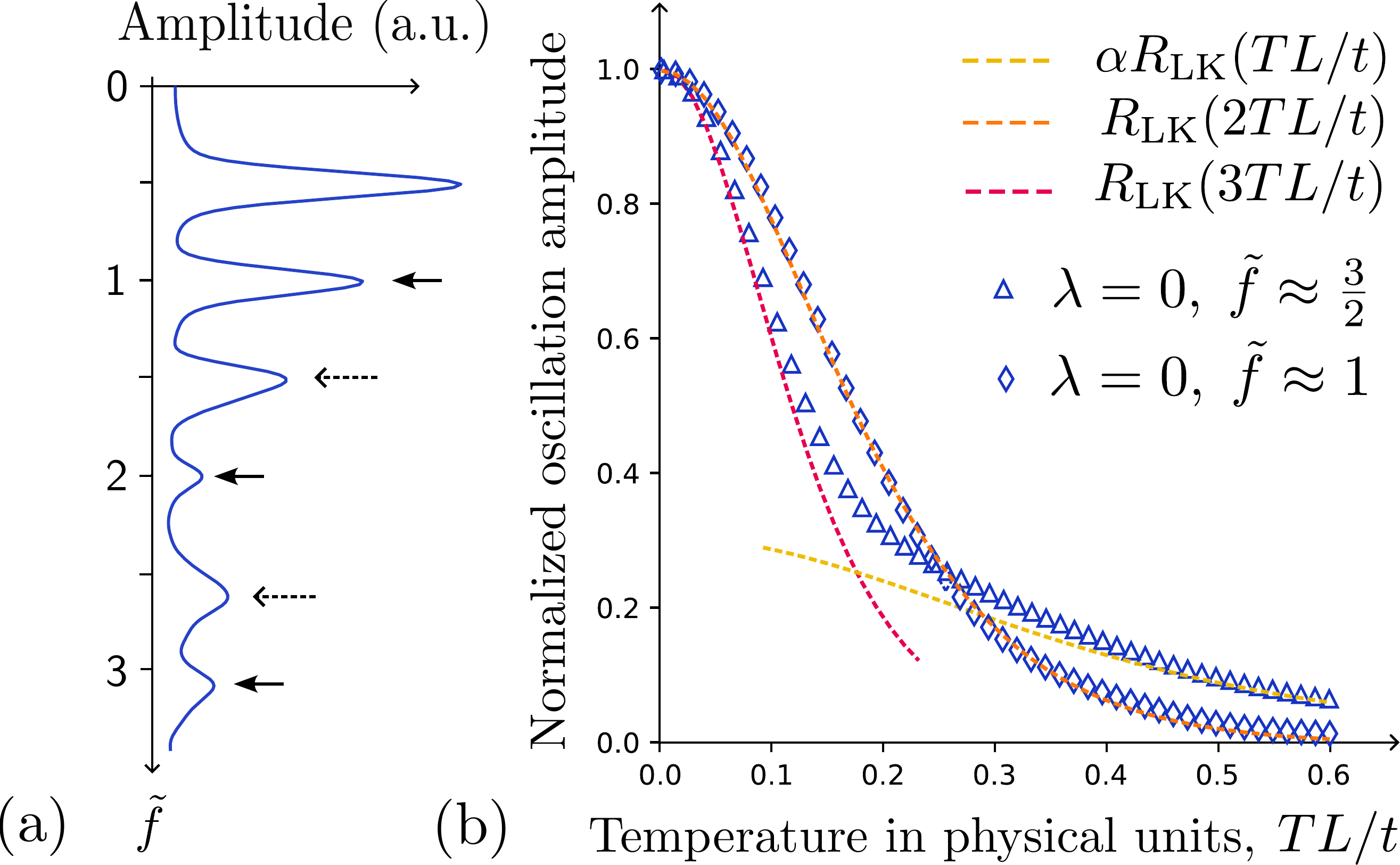}
\caption{\textbf{(a)} Fourier spectrum of quantum SO and their harmonics at $\lambda=0$ and $\delta_\varphi = 0.5$, in the limit $T \rightarrow 0, R_{\rm D}\rightarrow 1$, as a function of $\tilde f$. Arrows indicate peaks that arise partly (dashed) or solely (solid) from harmonics. 
  \textbf{(b)} Temperature dependence of the amplitude (for $\delta_\varphi = 0, \, R_{\rm D} \rightarrow 1$) of the $\tilde f=1$ and $\tilde f = 3/2$ peaks, compared with $R_{\rm LK}$ functions (dashed lines). Amplitudes are normalized to 1 at $T=0$, and we fitted $\alpha = 0.3$.}
\label{fig:harmonics}
\end{figure}

In Figure \ref{fig:harmonics}, we plot the amplitude of the peaks at $\tilde f = 3/2$, which is both $\tilde f_1$ and the third harmonic $3\tilde f_0$, and at $\tilde f = 1$, which only consists of the second harmonic $2\tilde f_0$. While the latter follows a dependence $R_{\rm LK}(2LT/t)$, the former is a sum of two contributions with distinct $T$ dependences $R_{\rm LK}(3LT/t)$ and $R_{\rm LK}(LT/t)$.\\

\section{Experimental aspects and phenomenology}
\label{sec:exper-cons}

\subsection{Quantum Sondheimer vs classical Sondheimer}

For trivial quadratic bands, namely $\tilde f =1/2$ at $\lambda=0$, we find that the fundamental frequency of quantum SO matches the usual frequency of classical SO. However, in general we found that quantum SO spectra have distinctive features: they (i) can show several peaks, and (ii) these can be located at any real value
-- for instance there are peaks at $\tilde f = \sqrt 2$, $\tilde f = \sqrt 6$, etc, for non-trivial Berry phase at $\lambda =1$. These two features are absent in classical SO. 
Here we briefly review the elementary physical pictures of both mechanisms.

The physical picture of quantum SO is completely straightforward and is summarized in Fig.\ref{fig:bigfigure}(c).
Each Landau level $n$, appearing as a dispersive band, is shifted vertically by $E_n\propto B$ when the magnetic field is varied.
This makes discrete energy levels, that near the middle of the band are evenly spaced by $\Delta_k\epsilon = \pi t/L = v_z \Delta k_z$,
periodically cross the Fermi level. This periodicity is fixed by the condition $E_n / \Delta_k\epsilon \in \mathbb Z$,
and there is one such condition for each Landau level $n$, whence several frequencies $\tilde f$ appearing in Figs.\ref{fig:bigfigure}(a),\ref{fig:damping}(a).
In experimental quantities, the amplitude of each peak, proportional to the LL degeneracy $N_{\sf \Phi}$, scales inversely with the total number of LLs involved:
quantum SO require this number to be small, otherwise the physics is that of classical SO. 

The physical picture of classical SO is in fact more complicated than the quantum case, and was pedagogically exposed in Ref.\cite{Guo_2026}.
To compensate for the smallness of each LL's contribution, one must enforce a matching condition that many discretized energy levels belonging to different LLs should cross the Fermi level simultaneously.
This is only possible when LL as well are evenly spaced, classically by $\Delta_n\epsilon = \omega_{\rm c}$.
Then, naively, said constraint would lead to the condition $\Delta_n\epsilon / \Delta_k\epsilon \in^? \mathbb Z$,
but a refinement due to the conservation of electron \emph{density} (instead of chemical potential), first noticed in Ref.\cite{Guo_2026}, modifies this into $\Delta_n\epsilon / \Delta_k\epsilon \in 2\mathbb Z$.
This correctly predicts the usual frequency of classical SO, which becomes the only frequency appearing in the spectrum of SO when the number of LLs involved is large.
How this ``resonance'' picture for classical SO relates to other derivations that do not involve any $k_z$ quantization \cite{nikolaenko2025theorysondheimermagnetooscillationssemiclassical}, in the limit of small magnetic fields, remains to be understood. In particular, the role of different boundary conditions (specular, diffusive, etc), and of relaxing the assumption of a mean free path much larger than the film thickness, remain to be clarified.

\subsection{Experimental status in few-LL settings}
\label{sec:exp-status-few}

Notably, oscillatory signatures of $k_z$ quantization \cite{yin2019dimensional} and $B$-periodic oscillations \cite{PhysRevB.108.235411} have recently been measured in graphite thin films in the quantum limit of large fields, with only two LLs involved,
that correspond to the lowest-lying levels at $\lambda=0$ in our model, $E^\pm_{0,0} = \pm \omega_{\rm c}/2$. 

In Ref.\cite{PhysRevB.108.235411} these are discussed as classical SO, as this work calls upon the semi-classical matching condition $\Delta_n\epsilon / \Delta_k\epsilon \in^? \mathbb Z$. Yet crucially, there is no physical need for such a matching condition when there are few LLs involved, each with a large degeneracy $N_{\sf \Phi}$ that yields sizeable quantum SO at its own frequency $f$. Besides, the argument of Ref.\cite{PhysRevB.108.235411} omits an important factor of 2 due to chemical potential oscillations \cite{Guo_2026}. 
We attribute the fact they still find agreement with the experimental data to their also considering a momentum step $2\Delta k_z=2\pi/{\sf a}L$ instead of $\Delta k_z$, whence a compensation of factors of 2.

Instead, we argue these are genuine \emph{quantum} SO, as we described here, and the correct picture is that of Fig.\ref{fig:bigfigure}(c). This is also supported by the fact that the phenomenon occurs at higher magnetic fields (approaching the quantum limit) than the SdH oscillations, whereas classical SO instead usually occur at lower fields than $1/B$ oscillations (see e.g.~Ref.~\cite{PhysRevB.84.125144}). 
A smoking-gun signature would be the presence of a superposition of several SO frequencies: one should look for these in graphite thin films under conditions where other LLs than $E^\pm_{0,0}$ are present.

\subsection{Summary of general features}
\label{sec:summary-general-f}

Here we briefly recapitulate some similarities and differences between classical Sondheimer (cSO), Shubnikov-de Haas (SdH), and quantum Sondheimer (qSO) oscillations : see Table \ref{tab:tab-summary}. 

\begin{table}[htbp]
  \centering
  \begin{tabular}{c|c|c|c}
    \hline\hline
   \vphantom{ \Big | } 
Type  & \,  cSO  \, & \,  SdH \, &\, qSO \, \\
    \hline
    \vphantom{ \Big | } 
  Periodicity & $L\omega_{\rm c}/t$ & \, $2\pi (\mu\pm t)/\omega_{\rm c}$  \, & \, $2\tilde f_n L\omega_{\rm c}/t\,\,\forall n$ \, \\
    \hline
     \vphantom{ \Big | } 
    Disorder  & \, $e^{-\Gamma L/(v/\sf a)}$ \cite{nikolaenko2025theorysondheimermagnetooscillationssemiclassical} \, & \, $e^{-2\pi \Gamma / \omega_{\rm c}}$ \, &  \, $e^{-2\Gamma L/t}$ \, \\
    \hline
     \vphantom{ \Big | } 
   Thermal & \, $R_{\rm LK}(\tfrac 1 4 \tfrac{ LT}{v/\sf a} \tfrac {\omega_{\rm c}} \mu)$ \cite{nikolaenko2025theorysondheimermagnetooscillationssemiclassical}\, & \, $R_{\rm LK}(\pi T/\omega_{\rm c})$ \, &  \, $R_{\rm LK}(LT/t)$\, \\
     \hline
     \vphantom{ \Big | } 
    Surface & \, $(1-r)^2$ \, & \, $1$ \, &  \, $e^{-\delta_\varphi^2/2}$\, \\
    \hline\hline
  \end{tabular}
  \caption{Short summary of some properties of cSO, SdH and qSO magneto-oscillations: periodicity, disorder decay (Dingle-like) factor, thermal damping factor, surface roughness related effects ($r \in [0,1]$ is the boundary reflection coefficient). Here $v/{\sf a}=\sqrt{2\mu t}$ and we recall other notations: number of layers $L$, interlayer spacing $\sf a$, tunneling $t$, cyclotron frequency $\omega_{\rm c}$, Fermi energy $\mu$, temperature $T$, disorder scattering rate $\Gamma$, standard dephasing $\delta_\varphi$ from surface roughness, and $\tilde f_n = E_n/\omega_{\rm c}$ for each LL energy $E_n$, $n\in \mathbb Z$.}
  \label{tab:tab-summary}
\end{table}

Experimentally relevant facts are:
\begin{itemize}
    \item[-] While SdH oscillations are periodic in $1/B$, SO (both classical and quantum) are periodic in $B$.
    \item[-] The Dingle factor suppresses SdH oscillations at small fields, while SO are favored by thin films and large mean free paths independent of $B$.
    \item[-] Thermal damping is absent at large fields for SdH oscillations, at small fields for classical SO, and is field-independent for quantum SO.
    \item[-] Classical SO vanish in the specular limit, whereas quantum SO are destroyed by surface roughness.
\end{itemize}

We note that qSO are absent from the ``summary'' Eq.\ 83 of Ref.\cite{nikolaenko2025theorysondheimermagnetooscillationssemiclassical} -- they would be the $(s=0,\, p=1)$ mode in their notations. The case of a few LLs is briefly mentioned, assuming a trivial quadratic dispersion in all three $(k_x,k_y,k_z)$ directions: here, we investigate thoroughly qSO for the general model Eq.\ref{eq:24}.

\subsection{Other potential smoking-gun criteria}

Looking further, several criteria can allow one to distinguish between quantum and classical SO,
beside the ones already mentioned regarding the number and positions of peaks in ${\rm FT}[\tilde \sigma_{xx}](f)$.
Notably, quantum SO should still exist even in the ``ultra-quantum'' limit where a single LL is involved (with caveats at fixed charge density -- see below).
In addition, their mechanism solely involves energy levels crossing the chemical potential, and so quantum SO should also be visible in \emph{thermodynamical} signatures:
the density of states $ \rho(\mu) = - (1/\pi V) {\rm Tr}\,{\rm Im}\,G(\mu) $ also has such oscillations, as its leading harmonics are contained in
\begin{align}
  \label{eq:7}
   \tilde \rho (\mu)   &= \frac {1}{\pi l_B^2} \left ( \frac{L}{\pi t} \right ) \frac 1 {{\sf a}L} \sum_n  
                           {\rm Re} \left ( e^{i2\pi  k_n^\star } [\sin(\pi k^\star_n /L)]^{-1} \right ) ,
\end{align}
for a detailed calculation see appendix \ref{sec:quant-sondh-oscill-thermo}. 
This has the same frequencies and the same damping factors $R_{\rm D}$, $R_{\rm LK}(TL/t)$ and $R_\Sigma$ as $\tilde \sigma_{xx}$.
While oscillations of thermodynamics properties are notoriously harder to measure in thin films than transport oscillations, this is still accessible to magnetic torque cantilever techniques~\cite{schwarz2002sawtoothlike},
and would provide yet another unambiguous signature of genuine quantum SO.

\subsection{Effect of a fixed carrier density}
\label{sec:effect-fixed-density}

\begin{figure}[!t]
\centering
\includegraphics[width=\columnwidth]{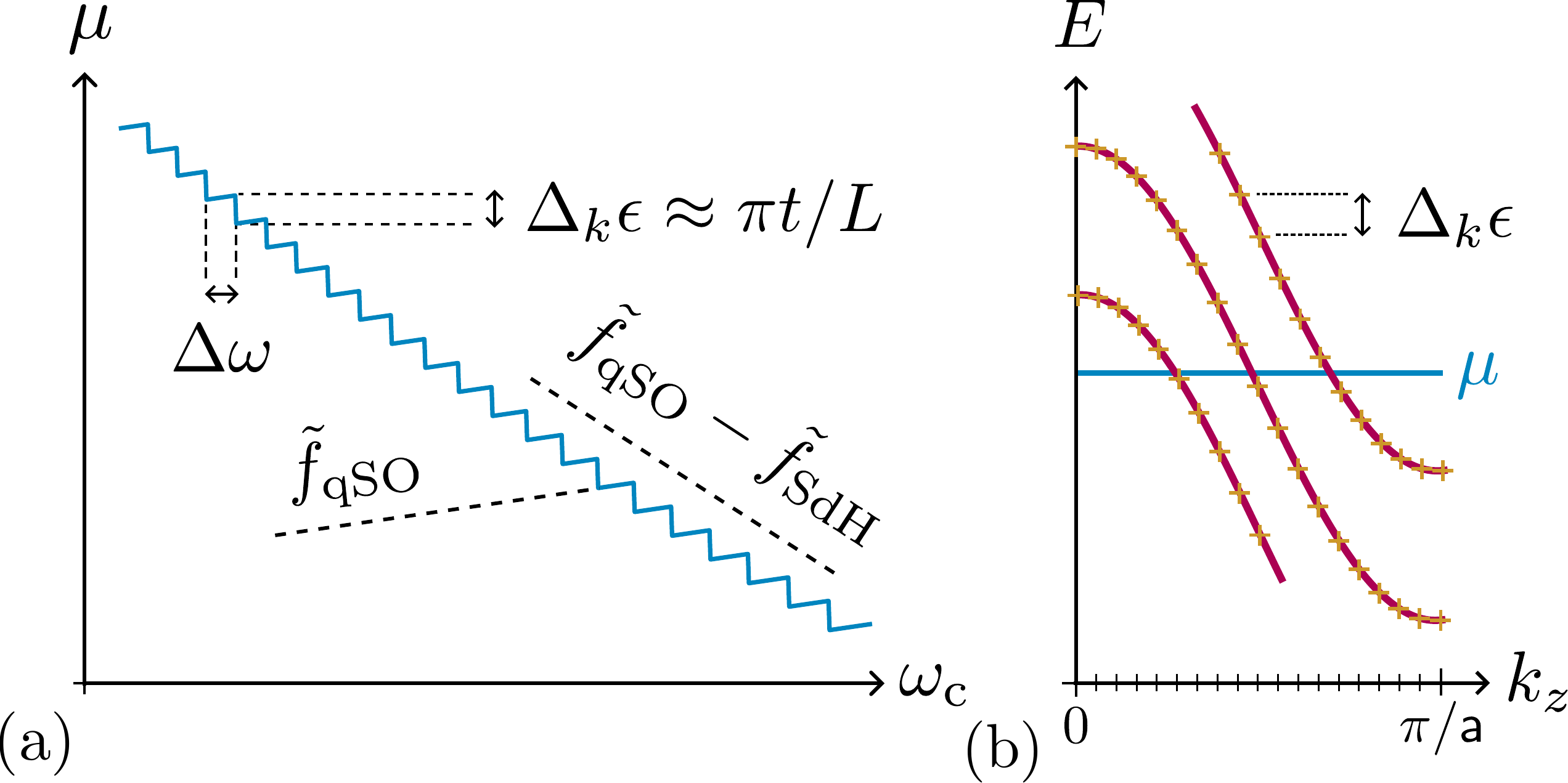}
\caption{\textbf{(a)} Schematic dependence of the chemical potential on the cyclotron frequency. While the small, SdH-like, sawtooth-shaped oscillations do not alter qSO frequencies, the slower evolution shifts all qSO peaks by a fixed amount, $\tilde f_n \mapsto \tilde f_{\rm eff}(n)$ given by Eq.\eqref{eq:tildef_eff}. \textbf{(b)} Schematic energy level distribution with respect to discrete $k_z$ (here for $\sf n = 3$).}
\label{fig:fixed-density}
\end{figure}

In 2D or quasi-2D systems without charge reservoirs, the chemical potential $\mu$ is generally not fixed, but instead the carrier density $\nu$ is constant. Then, because each energy sublevel's degeneracy per unit area $d=B/2\pi L$ is proportional to $B$, the chemical potential itself evolves with $B$: an instance of its dependence $\mu(B)$ in the case of linear $E(k_z)$ dispersions at the Fermi level is sketched in Fig. \ref{fig:fixed-density}.
This entails small SdH-type ($1/B$) oscillations of $\mu$, with local quasi-period $\Delta \omega \simeq \omega_{\rm c}^2/FL$, where $F=2\pi \nu/{\sf e}$ is the SdH oscillation frequency. 
This fixes a slope $\tilde f_{\rm SdH}=\Delta_k\epsilon/\Delta\omega = \pi t F/\omega_{\rm c}^2$.
In addition, each ``stair step'' has a slope $\tilde f_n$ for one of the $\sf n$ LLs crossing the Fermi level, $n \in \{n_1,\cdots,n_{\sf n}\}$. Hence the average stair step's slope is  $\tilde f_{\rm qSO} = \frac 1 {\sf n}\sum_{i=1}^{\sf n}\tilde f_{n_i}$.
Therefore, the net average slope of $\mu(\omega_{\rm c})$ is $\tilde f_{\rm qSO} - \tilde f_{\rm SdH}$. 

These fast chemical potential oscillation are relatively easy to distinguish from qSO spectra as they are $1/B$-periodic \cite{xiang2018thickness}, and are affected by various damping mechanisms. However, the slower evolution analyzed above, that survives damping, has crucial consequences. Indeed, the relative velocity (in Joules per Tesla), with respect to the Fermi level, of the discrete energy sublevels belonging to one LL $n$ is no longer $\tilde f_n = E_n/\omega_{\rm c}$ but 
\begin{equation}
\label{eq:tildef_eff}
    \tilde f_{\rm eff}(n) = \tilde f_n - \tilde f_{\rm qSO} + \tilde f_{\rm SdH} ,
\end{equation}
so the location of the corresponding peak in the FT is shifted by a fixed amount. 
We note because the $k_z$ dispersion is not linear over large energy scales, the varying slope $\tilde f_{\rm SdH}$ can result in smearing out the qSO frequencies (while this nonlinearity effect is essentially absent at fixed $\mu$, see Appendix \ref{sec:techn-deta-summ}). 

To overcome these spurious effects (that dominate the finite-size physics in Ref.\cite{xiang2018thickness} in $\rm WTe_2$), one way is to maintain $\mu$ fixed, so that $\tilde f_{\rm qSO} = 0 = \tilde f_{\rm SdH}$. This is necessarily the case for instance in a particle-hole symmetric system at charge neutrality (see Ref.\cite{PhysRevB.108.235411} in graphite which exhibits qSO without spurious SdH effects). This is also approximately the case when investigating small pockets where $\mu$ is maintained approximately constant by the presence of a much larger pocket acting as a reservoir. In the limit case of a single LL $n$ with no charge reservoir, $\tilde f_{n} = \tilde f_{\rm qSO}$ cancel against each other and Eq.\eqref{eq:tildef_eff} recovers the physics of plain SdH oscillations.

\section{Conclusion}
\label{sec:conclusion}

In this paper, we have derived a theory for quantum Sondheimer oscillations, both of transport and of thermodynamic properties, in conducting thin films. Our derivation holds for any intra-layer Hamiltonian with a simple tunnel coupling between layers. Using analytical calculations, we show that quantum SO, found in the limit of large $B$ fields, are a direct consequence of discretization of momentum in the transverse direction and Landau level quantization.
Our prediction for quantum SO shares similarities with classical SO, for instance their damping factors due to spectral and thermal broadening are similar. It also shows several differences, for instance in surface roughness effects, and crucially quantum SO happen at large magnetic fields whereas classical SO are generally found at low fields.

Our main finding, and a key feature of quantum SO, is that in the quantum limit, several systems of oscillations periodic in $B$ are present, one for each of the few LL involved. Each frequency is directly proportional to the energy of the corresponding LL. Thus, quantum SO allow to extract important information, such as the topological properties of band touching points, directly from the period of oscillations -- 
as opposed to their phase, as is the case for conventional $1/B$ magneto-oscillations. Whether topological properties could also be extracted from classical SO, and more generally how the quantum SO regime ($t/\omega_{\rm c}\sim 1$) crosses over to the classical one ($t/\omega_{\rm c}\gg 1$) in the limit of small $B$ fields and depending on the mean free path and boundary conditions, remain open theoretical questions.\footnote{These may have implications for recent experiments on cadmium thin films \cite{Guo_2026}, where the quantum-classical crossover is especially fuzzy. A possibly related experimental mystery is the absence of any reported SO in elemental bismuth.}

Although SO measurements were historically done in the classical (small field) limit, in particular in elemental metal thin films, novel experimental platforms and much larger magnetic fields available now open new possibilities. Beyond graphite thin films \cite{PhysRevB.108.235411}, natural avenues to explore include various semiconductor heterostructures. In particular, in transition metal dichalcogenides, large spin-orbit effects are associated with instances of nontrivial band topology, and high tunability can be achieved. Overall, by providing a direct signature of nontrivial topology, quantum SO seem a promising avenue for new discoveries. 

On the theory side, much remains to be investigated. First, as we already mentioned, a quantitative understanding, e.g.\ based on unbiased lattice numerics, of the quantum-classical SO crossover and the role of surface disorder, is currently missing. Then, other models for the 3D dispersion than our Eq.\eqref{eq:24} (which is a tight-binding stack of 2D systems) should also be explored: for instance, how about thin films of 3D topological band touchings, e.g. Weyl semimetals? This could also help understand the $B$-field dependence of oscillation amplitudes, and its dependence on FS geometry \cite{Guo_2026}. Looking even further, one promising aspect of quantum SO is that the energy discretization is provided by the finite slab thickness, instead of the traditional quantization of orbits around the Fermi surface: thus it seems possible to look for oscillations in systems without an electron Fermi surface, for instance inverted insulators \cite{PhysRevLett.115.146401} or systems with emergent fractionalized fermions \cite{bonetti2024quantum} where mechanisms for magneto-oscillations have been proposed.\\

\section*{Acknowledgements}

We thank Pavel Nosov and Alexander Nikolaenko for a useful discussion, and Jean-Noël Fuchs and Kamran Behnia for helpful comments on our manuscript.
We acknowledge support from the Imperial-TUM flagship partnership, from the Deutsche Forschungsgemeinschaft (DFG, German Research Foundation) under Germany’s Excellence Strategy–EXC– 2111–390814868, DFG grants No. KN1254/1-2, KN1254/2-1, and TRR 360 - 492547816, as well as the Munich Quantum Valley, which is supported by the Bavarian state government with funds from the Hightech Agenda Bayern Plus.

\bibliography{library}

\appendix

\section{Band geometrical properties}
\label{sec:few-notes-about}

Here we quickly recall the quantum geometry of bands in zero field.
One can write $ H_\lambda = \mb d(\bs k)\cdot \bs \sigma$ where
\begin{subequations}
  \begin{align}
  d_x &= \lambda\, (k_x^2-k_y^2)/2m ,\\
  d_y &=  \lambda\, 2k_xk_y/2m ,\\
  d_z&= (1-\lambda )\, k^2 /2m .
\end{align}
\end{subequations}
In other words, in polar coordinates where $(k_x,k_y)=\sqrt{k^2-k_z^2}\,\big (\cos(\phi),\sin(\phi)\big )$, 
\begin{align}
   {\mb d}(\phi) &= \frac{k^2}{2m}\big ( \lambda \cos(2\phi),\lambda \sin(2\phi), 1-\lambda \big ) .
\end{align}
From this the energies are
\begin{align}
  E_\pm(\bs k) &= \pm \sqrt{\lambda^2 + (1-\lambda^2)}\,k^2/2m .
\end{align}

The topology depends on the normalized vector
\begin{align}
  \label{eq:44}
  \hat {\mb d}(\phi) = \tfrac 1 {\sqrt{\lambda^2 + (1-\lambda)^2 }} \big ( \lambda \cos(2\phi),\lambda \sin(2\phi), 1-\lambda \big ) ,
\end{align}
which in spherical coordinates reads
\begin{align}
  \label{eq:46}
  \hat  {\mb d} = \big (\sin\theta \cos\varphi,\sin\theta \sin\varphi,\cos\theta),
\end{align}
where $\varphi(\phi)=2\phi$ and 
\begin{align}
  \label{eq:47}
  \cos\theta = \frac{1-\lambda} {\sqrt{\lambda^2+(1-\lambda)^2}},
\;
\sin\theta = \frac{\lambda} {\sqrt{\lambda^2+(1-\lambda)^2}}.
\end{align}
Here $\theta$ depends only on $\lambda$.
One can readily identify the winding number $w_\pm = \mp 2$.

The eigenstates can be written as
\begin{align}
  \label{eq:49}
  \begin{pmatrix}
    |u_+\rangle \\ |u_-\rangle
  \end{pmatrix} =
  \begin{bmatrix}
    \cos\frac{\theta}{2} & \sin\frac{\theta}{2} \\ -\sin\frac{\theta}{2} & \cos\frac{\theta}{2}
  \end{bmatrix}
                                                                           \begin{pmatrix}
                                                                             1 |\!\uparrow\rangle \\ e^{i\varphi}|\!\downarrow\rangle
                                                                           \end{pmatrix} .
\end{align}
The Berry connection is
\begin{align}
  \label{eq:50}
  A_\phi^\pm =  i\langle u_\pm|\partial_\phi u_\pm \rangle = \mp (1-\cos \theta ).
\end{align}
The Berry phase accumulated around a loop encircling the origin is
\begin{align}
  \label{eq:51}
  \gamma_{\pm} &= \int_0^{2\pi} A^\pm_\phi\,\text d\phi = \mp \,2\pi  (1-\cos \theta ) ,
\end{align}
that is to say explicitly for the lower band
\begin{align}
  \label{eq:52}
  \gamma(\lambda) &=  2\pi \left( 1- \frac{1-\lambda} {\sqrt{\lambda^2+(1-\lambda)^2}} \right).
\end{align}
One can check that in the trivial limit the Berry phase vanishes, $\gamma(0)=0$, and in the other limit $\gamma(1)=2\pi$.

  \section{Landau level energies and eigenstates}
  \label{sec:land-levels-energ}

  First we consider the single-layer problem.
  We use the Landau gauge, where $p_x = -i\partial_x - y/l_B^2$ and $p_y = -i \partial_y$, where $\ell_B=({\sf e}B)^{-1/2}$.
  Then the eigenfunctions are
  \begin{align}
    \label{eq:23}
    \psi_{\tilde n,{\rm p}_x}(\bs r) &=  L_x^{-1/2} {e}^{i {\rm p}_x x} \psi_{\tilde n} \big (y - {\rm p}_x \big ) ,
  \end{align}
  where
  \begin{align}
    \label{eq:57}
    \psi_{\tilde n}(\xi) = \frac{\pi^{-\nicefrac 14}}{\sqrt{2^{\tilde n} \tilde n!}} e^{- \xi ^2 /2 } H_{\tilde n} \left [ \xi \right ] 
  \end{align}
  and $H_{\tilde n}$ is the $\tilde n$th Hermite polynomial, normalized such that $\int \text d \xi \,\psi_{\tilde n'}(\xi) \psi_{\tilde n}(\xi)=\delta_{\tilde n,\tilde n'}$.
  We recall the ladder operators
  \begin{subequations}
    \begin{align}
  \label{eq:45}
  a^\dagger &=  \tfrac 1 {\sqrt 2} \left ( p_x +  i p_y \right ) ,\\
  a &= \tfrac 1 {\sqrt 2} \left (  p_x - i p_y  \right ),
\end{align}
  \end{subequations}
  that verify
  \begin{subequations}
    \begin{align}
  \label{eq:42}
  a^\dagger  (-1)^{\tilde n} \psi_{\tilde n, {\rm p}_x} &= \sqrt{\tilde n+1} (-1)^{\tilde n+1} \psi_{\tilde n+1, {\rm p}_x} ,\\
  a (-1)^{\tilde n} \psi_{\tilde n, {\rm p}_x} &= \sqrt{\tilde n} (-1)^{\tilde n-1} \psi_{\tilde n-1, {\rm p}_x} .
\end{align}
  \end{subequations}
  From now on we use the notation $|\tilde n\rangle \equiv (-1)^{\tilde n}\psi_{\tilde n}$, keeping implicit the ${\rm p}_x$ degeneracy.

With the replacements
\begin{align}
  \label{eq:15}
  p_x = (a + a^\dagger)/\sqrt 2 , \; p_y = (a - a^\dagger)/i\sqrt 2 ,
\end{align}
the Hamiltonian takes the following form:
\begin{align}
  \label{eq:2}
  H_\lambda = \omega_{\rm c}
  \begin{bmatrix}
    (1-\lambda) \big ( a^\dagger a + \tfrac 1 2 \big  )& \lambda (a^\dagger)^2 \\
    \lambda a^2 & -(1-\lambda)  \big ( a^\dagger a + \tfrac 1 2 \big)
  \end{bmatrix} .
\end{align}

In the sectors $\tilde n\geq 2$, one can work in the subspace $\left \{ |\tilde n,\uparrow\rangle; |\tilde n-2,\downarrow\rangle \right \}$ where 
\begin{align}
  H_{\tilde n}(\lambda) =
 \omega_{\rm c} \begin{bmatrix}
    (1-\lambda)\left(\tilde n+\tfrac12\right)&\lambda\,\sqrt{\tilde n(\tilde n-1)}
\\ \lambda\,\sqrt{\tilde n(\tilde n-1)}&-(1-\lambda)\left(\tilde n-\tfrac32\right)
\end{bmatrix},
\end{align}
with eigenvalues
\begin{align}
  \label{eq:9}
  E_{\tilde n}^\pm(\lambda) &=  \omega_{\rm c} (1-\lambda) \\
  &\nonumber  \pm\; \omega_{\rm c} 
\sqrt{(1-\lambda)^2 (\tilde n - \tfrac 1 2)^2 + \lambda^2 \tilde n(\tilde n-1)}
\end{align}
and associated eigenvectors
\begin{subequations}
  \begin{align}
  \label{eq:10}
 |\tilde n,+\rangle &= \cos\frac{\theta_{\tilde n}}{2}\,|\tilde n,\uparrow\rangle + \sin\frac{\theta_{\tilde n}}{2}\,|\tilde n-2,\downarrow\rangle ,\\
  |\tilde n,-\rangle &= -\sin\frac{\theta_{\tilde n}}{2}\,|\tilde n,\uparrow\rangle + \cos\frac{\theta_{\tilde n}}{2}\,|\tilde n-2,\downarrow\rangle ,
\end{align}
\end{subequations}
where
\begin{align}
  \label{eq:11}
  \tan\theta_{\tilde n} = \frac{\lambda \sqrt{\tilde n(\tilde n-1)}}{(1-\lambda)(\tilde n-\tfrac 1 2)} .
\end{align}
Explicitly, defining
\begin{align}
  \label{eq:5}
  \rho_{\tilde n} = \sqrt{(1-\lambda)^2 (\tilde n-\tfrac 1 2)^2+\lambda^2 \tilde n(\tilde n-1)},
\end{align}
one has
\begin{subequations}
  \begin{align}
  \label{eq:33}
    \cos(\theta_{\tilde n}/2) &= \sqrt{(\rho_{\tilde n} + (1-\lambda)(\tilde n-\tfrac 1 2))/2\rho_{\tilde n}},\\
    \sin(\theta_{\tilde n}/2) &= \sqrt{(\rho_{\tilde n} - (1-\lambda)(\tilde n-\tfrac 1 2))/2\rho_{\tilde n}} .
\end{align}
\end{subequations}

Written compactly,
\begin{align}
  \label{eq:16}
 |\tilde n,s\rangle &= R^{(\tilde n)}_{s\sigma} |\tilde n-1+\sigma,\sigma\rangle , \quad \tilde n\geq 2 , \sigma=\uparrow\downarrow,
\end{align}
where
\begin{align}
  \label{eq:58}
 R^{(\tilde n)} &=   \begin{bmatrix}
    \cos(\theta_{\tilde n}/2) & \sin(\theta_{\tilde n}/2) \\ - \sin(\theta_{\tilde n}/2) & \cos(\theta_{\tilde n}/2) 
\end{bmatrix}.
\end{align}

In addition, there are two uncoupled modes, 
\begin{subequations}
  \begin{align}
  \label{eq:12}
    E_0^\uparrow &= +\tfrac 1 2 \omega_{\rm c}(1-\lambda) ,\\
    E_1^\uparrow &= +\tfrac 3 2 \omega_{\rm c}(1-\lambda) ,
\end{align}
\end{subequations}
that are smoothly connected to topologically protected zero-modes in the limit $\lambda=1$.

\section{Velocity operator}
\label{sec:velocity-operator}

The velocity operator has two contributions,
\begin{align}
  \label{eq:14}
  \hat v_x &= \hat v_x^{\rm orb} + \hat v_x^{\rm od},\\
  \hat v_x^{\rm orb} &= \omega_{\rm c}\frac{1-\lambda}{\sqrt 2} (a + a^\dagger) \sigma^z \\
  \hat v_x^{\rm od} &=\omega_{\rm c}\lambda \sqrt 2
  \begin{bmatrix}
    0 & a^\dagger \\ a & 0
  \end{bmatrix} .
\end{align}
The orbital part is
\begin{align}
  \label{eq:17}
  \langle \tilde n,\sigma|  \hat v_x^{\rm orb} | \tilde n',\sigma'\rangle
  &= \delta_{\sigma,\sigma'}\sigma \omega_{\rm c} (1-\lambda)/\sqrt 2 \nonumber \\
  &\times (\sqrt{\tilde n'}\delta_{\tilde n',\tilde n+1} + \sqrt{\tilde n}\delta_{\tilde n',\tilde n-1}),
\end{align}
well-defined for all values of $\tilde n\geq 0,\tilde n'\geq 0$.
The off-diagonal part is
\begin{align}
  \label{eq:19}
  \langle \tilde n,\sigma|  \hat v_x^{\rm od} | \tilde n',\sigma'\rangle
  &= \omega_{\rm c} \lambda \sqrt 2 (1-\delta_{\sigma,\sigma'}) \sqrt{{\rm max}(\tilde n,\tilde n')} \nonumber \\
  &\times (\delta_{\tilde n,\tilde n'-\sigma'} = \delta_{\tilde n-\sigma,\tilde n'}) ,
\end{align}
also well-defined for all $\tilde n\geq 0, \tilde n'\geq 0$.
Then for matrix elements between generic states $(\tilde n\geq 2,\tilde n'\geq 2)$,
\begin{align}
  \label{eq:18}
  &  \langle \tilde n,s|  \hat v_x | \tilde n',s'\rangle \\
  &= R^{(\tilde n)}_{s\sigma} \langle \tilde n-1+\sigma,\sigma| \hat v_x| \tilde n'-1+\sigma',\sigma'\rangle R^{(\tilde n')}_{s'\sigma'} . \nonumber
\end{align}
Because one assumes $\tilde n\geq 2,\tilde n'\geq 2$, necessarily $\tilde n-1+\sigma\geq 0,\tilde n'-1+\sigma'\geq 0$ so all matrix elements are well-defined.
As for matrix elements that involve $|\eta,\uparrow\rangle$ together with a generic state, for $\eta \in \{0,1\}$, 
\begin{align}
  \label{eq:21}
 \langle \tilde n,s|  \hat v_x | \eta,\uparrow \rangle
  &= R^{(\tilde n)}_{s\sigma} \langle \tilde n-1+\sigma,\sigma|  \hat v_x | \eta,\uparrow \rangle .
\end{align}
Because one assumes $\tilde n\geq 2$, necessarily $\tilde n-1+\sigma\geq 0$ so all matrix elements are well-defined.
Finally, of course $ \langle \eta,\uparrow|  \hat v_x | \eta',\uparrow \rangle $ is well-defined for $\eta,\eta' \in \{0,1\}$
and is given by Eqs.\eqref{eq:17},\eqref{eq:19}.

These can now be inserted into the conductivity kernel,
\begin{align}
  \label{eq:26}
  \sigma_{xx}(\mu) &= - \frac{\sf e^2}{\pi} \frac 1 {V } \sum_{{\rm p}_x} {\rm Tr}_{k,n}\left [ \hat v_x \, {\rm Im}G(\mu) \,\hat v_x {\rm Im}G(\mu) \right ]
\end{align}
which because $V = L_xL_y{\sf a}L $ and $\sum_{{\rm p}_x} =N_{\sf \Phi}$ is equal to Eq.\eqref{eq:27}.
Explicitly in terms of matrix elements of the eigenbasis,
\begin{align}
  \label{eq:25}
 & {\rm Tr}_{k,n}\left [ \hat v_x\, {\rm Im}G(\mu) \,\hat v_x\, {\rm Im}G(\mu) \right ] \nonumber \\
  &= \sum_{k=1}^L \sum_{n,n' } \big | \langle n | \hat v_x | n' \rangle \big |^2 \, {\rm Im}G_{n,k}(\mu)\,  {\rm Im}G_{n',k}(\mu) .  
\end{align}

\section{Analytical details}
\label{sec:techn-deta-summ}

Now we Poisson-resum over $k \in \llbracket 1, L \rrbracket$, using the formula
\begin{align}
  \label{eq:6}
  \sum_{k=1}^{L} {\rm g}(k) &= \sum_{r\in \mathbb Z} \int_{1/2}^{L+1/2}\text dk \, e^{i2\pi r k}{\rm g}(k) ,
\end{align}
applied to the sum over $k$ in Eq.\eqref{eq:25}. Focusing on the fundamental harmonics $r = \pm 1$, one defines
\begin{align}
  \label{eq:28}
  \tilde \sigma_{xx}(\mu) &= - \frac{\sf e^2}{2\pi^2} l_B^{-2}\, \sum_{n,n'}\big | \langle n | \hat v_x | n' \rangle \big |^2 \, \sum_{r =\pm } \\
                         &\times \int_{1/2}^{L+1/2}\text dk \, e^{i2\pi rk} \, {\rm Im}G_{n,k}(\mu)\,  {\rm Im}G_{n',k}(\mu) .  \nonumber 
\end{align}
Here, the integral bounds can actually be extended to infinity rigorously. Indeed, the integrand is $L$-periodic,
so it is possible to use contour integration provided one counts only the poles whose real part is within $[0,L]$.
To evidence the poles and their residues, rewrite to the first order
\begin{align}
  \label{eq:8}
  &  \mu - E_n + t \cos(\pi k/L) + i \Gamma_n \nonumber \\
  &\approx - (\pi t/L) \, (k- k^\star_n)\, \sin(\pi k^\star_n /L) ,
\end{align}
where as defined in the main text,
\begin{align}
  \label{eq:59}
  k^\star_{n} &=  L \arccos[ (E_n - \mu - i \Gamma_n)/t] /\pi .
\end{align}
Notice that $0 \leq {\rm Re}(k^\star_n) \leq L$.
We will use, and prove later, that $k^\star_n$ is in the upper half-plane, $ {\rm Im}(k^\star_n) >0$.
Then it is sufficient to do the integration for $r=+1$, since $r=-1$ simply yields the complex conjugate result.
A straightforward calculation produces the result diaplayed in the main text, Eq.\eqref{eq:54}.

Now we analyze the convergence of expression Eq.\eqref{eq:54}.
First note that the inverse sine factors diverge for  $\arccos[(E_n-\mu - i \Gamma_n)/t] $ going to $0$ or $\pi$:
we will see that this is compensated by factors becoming exponentially small in these limits.
Indeed, writing $\zeta_n = (E_n-\mu)/ t$ and looking at $|\zeta_n| \rightarrow 1$, one has the asymptotics
\begin{align}
  \label{eq:20}
  \arccos\left [ \zeta_n - i\Gamma_n/t \right ]
  &\approx {\rm sign}(\zeta_n)\,\sqrt{2(1-\left | \zeta_n \right |)} \nonumber \\
    &+ i (\Gamma_n/t)/\sqrt{2(1-\left | \zeta_n \right |)}.
\end{align}
The imaginary part is positive and diverges as $|\zeta_n| \rightarrow 1$,
which proves the statement that the complex exponential factor $e^{i 2\pi k_n^\star}$ decreases exponentially in that limit.
Therefore, the sum in Eq.\eqref{eq:54} gets contributions from moderate to small values of $|E_n-\mu|/ t$, for which instead
\begin{align}
  \label{eq:22}
  &\arccos\left [ (E_n-\mu)/t - i\Gamma_n/t \right ] \nonumber \\
  &\approx \pi/2 - (E_n-\mu)/t +  i\Gamma_n/t .
\end{align}
This leads to Eq.\eqref{eq:1} in the main text.

\section{SdH oscillations}
\label{sec:sdh}

The term $r=0$ in Eq.\eqref{eq:6} contains SdH oscillations: in this appendix we compute the latter explicitly. We start from 
\begin{align}
\overline\sigma_{xx}(\mu) &= - \frac{\sf e^2}{2\pi^2} l_B^{-2}\, \sum_{n,n'}\big | \langle n | \hat v_x | n' \rangle \big |^2 \\
  &\times \int_{0}^{L}\text dk \,  {\rm Im}G_{n,k}(\mu)\,  {\rm Im}G_{n',k}(\mu) .  \nonumber 
\end{align}
Within this appendix, we will focus on the simpler regime where a large number of Landau levels is involved, assuming $|\mu\pm t| \gg \omega_{\rm c}$ as well as $\mu \geq t >0$. Thus, we restore indices $n \rightarrow (\tilde n,\sigma)$, and oscillating physical quantities are dominated by large values of $\tilde n$, where the dispersion recovers a classical form,
\begin{align}
    E_{\tilde n}^\pm(\lambda) 
    &\approx \pm \,\tilde n \, \tilde\omega_{\rm c}^\lambda , \\
    \tilde\omega_{\rm c} ^\lambda &= \omega_{\rm c}\sqrt{(1-\lambda)^2+\lambda^2} .
\end{align}
Then, SdH oscillations will arise from Poisson resummation over $\tilde n$: namely,
\begin{equation}
     \sum_{\tilde n=0}^{\infty} {\sf g}(\tilde n) = \sum_{r' \in \mathbb Z} \int_{-1/2}^{\infty}\text d \tilde n \, e^{i2\pi r' \tilde n}{\sf g}(\tilde n) .
\end{equation}
Focusing on the leading SdH harmonics (that are lowest order in Dingle factors), we keep only the $r'=\pm$ terms. To evaluate the integral over $\text d \tilde n$ we identify the poles, 
\begin{equation}
 \tilde n^\star_k + i \Gamma/\tilde\omega_{\rm c}
 = [\mu + t \cos(\pi k/L) + i \Gamma ]/\tilde\omega_{\rm c} ^\lambda ,
\end{equation}
and perform contour integration, which results in
\begin{align}
\overline \sigma_{xx}(\mu) &= - \frac{\sf e^2}{\pi^2} l_B^{-2}\int_0^L \text dk  \,(2\pi \Gamma /\tilde\omega_{\rm c} ^\lambda) \, e^{-2\pi \Gamma/\tilde\omega_{\rm c} ^\lambda } \, {\cos} (2\pi  n^\star_k ) \nonumber \\
&\times \sum_{\tilde n,\sigma} \frac{\big | \langle n^\star_k , +| \hat v_x | \tilde n,\sigma \rangle \big |^2}
                     {[\mu - E_{\tilde n}^{\sigma} + t \cos(\pi k/L) ]^2+  \Gamma^2 }.
\end{align}
The integral over $\text dk$ is governed by the fast oscillating ${\cos} (2\pi  n^\star_k ) $ factor, and can be computed using the stationary phase approximation. This yields
\begin{align}
 \overline \sigma_{xx}(\mu)
  &= - \frac{\sf e^2}{2\pi^2} l_B^{-2} (2\pi \Gamma /\tilde\omega_{\rm c} ^\lambda) \, (L/\pi) \, e^{-2\pi \Gamma/\tilde\omega_{\rm c} ^\lambda }
    \sqrt{\tilde\omega_{\rm c} ^\lambda/t} 
    \nonumber \\
    &\times \sum_{\varsigma = \pm}
    {\cos}\big( 2\pi \tilde n^\star_{\lambda,\varsigma}  - \varsigma \tfrac \pi 4 \big ) 
    \sum_{\tilde n,\sigma} \frac{\big | \langle \tilde n^\star_{\lambda,\varsigma},\sigma  | \hat v_x | \tilde n,\sigma \rangle \big |^2} {[\mu - E_{\tilde n}^{\sigma} + \varsigma t ]^2+  \Gamma^2 },
\end{align}
where $\tilde n^\star_{\lambda,\varsigma}
= (\mu+\varsigma t)/\tilde\omega_{\rm c} ^\lambda $. Due to the cosine $k_z$ dispersion, there are two SdH oscillation frequencies, and a beating pattern emerges. This phenomenon is absent in the case of a parabolic $k_z$ dispersion, as studied in Ref.\cite{nikolaenko2025theorysondheimermagnetooscillationssemiclassical}. Ultimately this is a simple consequence of the position of van Hove points set by different $k_z$ dispersions.

\section{Quantum Sondheimer oscillations in thermodynamics}
\label{sec:quant-sondh-oscill-thermo}

Here we show that quantum SO can also, in principle, be found in thermodynamical quantities.
Many of these derive simply from the density of states, 
\begin{align}
  \label{eq:3}
  \rho(\epsilon) &= - \frac 1 \pi \frac 1 {V} \sum_{{\rm p}_x}{\rm Tr}_{k,n}{\rm Im}\,G(\epsilon) \\
  &= - \frac {l_B^{-2}}{2\pi^2} \frac 1 {{\sf a}L} \sum_{n,k}{\rm Im} \left [ \epsilon - E_n + t \cos(k \pi/L) + i \Gamma_n \right ]^{-1} . \nonumber
\end{align}

We then perform the Poisson-resummation over $k$, using again Eq.\eqref{eq:6}, and computing the integral over $k$, 
\begin{align}
  \label{eq:4}
  \tilde \rho (\epsilon) &=  - \frac {l_B^{-2}}{2\pi^2} \frac 1 {{\sf a}L} \sum_n {\rm Im} \sum_{r =\pm } \\
  & \times \int_{1/2}^{L+1/2}\text dk
                          \, e^{i2\pi  r k } \left [ \epsilon - E_n + t \cos(k \pi/L) + i \Gamma_n \right ]^{-1} , \nonumber
\end{align}
with the same contour integration method. Because $k^\star_n$ is in the upper-half plane, only $r=+$ contributes,
which yields the result quoted in the main text, Eq.\eqref{eq:7}.

Using again the approximation Eq.\eqref{eq:22}, one finds again the same Dingle factor $R_{\rm D} = e^{-2L\Gamma/t}$,
thermal damping factor $R_{\rm LK}(TL/t)$ and surface smearing factor $R_\Sigma$ as for $\tilde \sigma_{xx}(\mu)$.

\end{document}